\documentclass[a4paper]{IEEEtran}
\usepackage{cite}
\ifCLASSINFOpdf
  \usepackage[pdftex]{graphicx}
\else
  \usepackage[dvips]{graphicx}
\fi
\usepackage[T1]{fontenc}
\usepackage[cmex10]{amsmath}
\usepackage{array}
\usepackage{theorem}
\usepackage[caption=false,font=footnotesize]{subfig}
\usepackage{fixltx2e}
\usepackage{url}
\usepackage{acronym}
\usepackage{comment}
\usepackage{soul,color}
\usepackage{multirow}

{\theorembodyfont{\rmfamily}
   }
{\theorembodyfont{\rmfamily}
   }
{\theorembodyfont{\rmfamily}
   }
{\theorembodyfont{\rmfamily}
   }

{\theorembodyfont{\rmfamily}
   }
{\theorembodyfont{\rmfamily}
   }

\acrodef{LDPC}{low-density parity-check}
\acrodef{MDPC}{moderate-density parity-check}
\acrodef{QC}{quasi-cyclic}
\acrodef{QC-LDPC}{quasi-cyclic low-density parity-check}
\acrodef{QC-MDPC}{quasi-cyclic moderate-density parity-check}
\acrodef{RSA}{Rivest, Shamir, Adleman}
\acrodef{BF}{bit flipping}
\acrodef{SPA}{sum product algorithm}
\acrodef{RDF}{random difference families}
\acrodef{ISDA}{information set decoding attacks}
\acrodef{DCA}{dual code attacks}
\acrodef{WF}{work factor}
\acrodef{BER}{bit error rate}
\acrodef{FER}{frame error rate}
\acrodef{CER}{codeword error rate}
\acrodef{BLER}{block error rate}
\acrodef{PLS}{Physical layer security}
\acrodef{SNR}{signal-to-noise ratio}
\acrodefplural{SNR}[SNRs]{signal-to-noise ratios}
\acrodef{eBCH}{extended Bose-Chaudhuri-Hocquenghem}
\acrodef{CCs}{convolutional codes}
\acrodef{CC}{convolutional code}
\acrodef{UB}{union bound}
\acrodef{TUB}{truncated union bound}
\acrodef{QSFC}{quasi-static fading channel}
\acrodefplural{QSFC}[QSFCs]{quasi-static fading channels}
\acrodef{FFC}{fast fading channel}
\acrodef{CSI}{channel state information}
\acrodef{AWGN}{additive white Gaussian noise}
\acrodef{NMS}{normalized min-sum}
\acrodef{LLR}{log likelihood ratio}
\acrodef{LLR-SPA}{log-likelihood ratio sum-product algorithm}
\acrodef{BCC}{broadcast channel with confidential messages}
\acrodef{UEP}{unequal error protection}
\acrodef{BPSK}{binary phase shift keying}
\acrodef{QAM}{quadrature amplitude modulation}
\acrodef{ML}{maximum likelihood}
\acrodef{AONT}{all-or-nothing transform}
\acrodefplural{AONT}[AONTs]{all-or-nothing transforms}
\acrodef{PC}{protection class}
\acrodefplural{PC}[PCs]{protection classes}
\acrodef{MIMO}{multiple-input multiple-output}
\acrodef{OSD}{ordered statistics decoding}
\acrodef{ISD}{information set decoding}

\def\gaveE{\bar{\gamma}^{(E)}}
\def\gaveB{\bar{\gamma}^{(B)}}
\def\gE{{\gamma}^{(E)}}
\def\gB{{\gamma}^{(B)}}

\begin{document}

\title{Practical LDPC coded modulation schemes for the fading broadcast channel with confidential messages
\thanks{This work was supported in part by the MIUR project ``ESCAPADE''
(Grant RBFR105NLC) under the ``FIRB -- Futuro in Ricerca 2010'' funding program.}
}

\author{\IEEEauthorblockN{Marco Baldi, Nicola Maturo, Giacomo Ricciutelli, Franco Chiaraluce,\\}
\IEEEauthorblockA{DII, Universit\`a Politecnica delle Marche,\\
Ancona, Italy\\
Email: \{m.baldi, n.maturo, f.chiaraluce\}@univpm.it}, g.ricciutelli@gmail.com}

\maketitle

\pagestyle{empty}
\thispagestyle{empty}

\begin{abstract}
The broadcast channel with confidential messages is a well studied scenario from the theoretical standpoint, but there is still lack of practical schemes able to
achieve some fixed level of reliability and security over such a channel.
In this paper, we consider a quasi-static fading channel in which both public and private messages must be sent from the
transmitter to the receivers, and we aim at designing suitable coding and modulation schemes to achieve such a target.
For this purpose, we adopt the error rate as a metric, by considering that reliability (security) is achieved when
a sufficiently low (high) error rate is experienced at the receiving side.
We show that some conditions exist on the system feasibility, and that some outage probability must be tolerated
to cope with the fading nature of the channel.
The proposed solution exploits low-density parity-check codes with unequal error protection, which are able to guarantee two different levels of 
protection against noise for the public and the private information, in conjunction with different modulation schemes for the public and the private message bits.
\end{abstract}


\section{Introduction}
\label{sec:Intro}

One of the basic transmission models for physical layer security is the \ac{BCC} \cite{Csiszar1978}.
In this model, there is one transmitter (Alice) who sends both broadcast and confidential information over the channel.
The authorized receiver (Bob) is able to decode the whole information, while the non-authorized receiver (Eve) can
only have access to the public information, but she should be unable to obtain the confidential information.
Bob's and Eve's channels are generally different one each other.
A practical context in which the \ac{BCC} model can be applied is the integration of multiple services at the physical layer. 
For example, a wireless network could provide a free broadcast service to all users, and also exploit the same channel to
provide another service which is restricted to a subset of users.

The \ac{BCC} model has been extensively studied from the information theory standpoint, mostly with the aim
of computing the secrecy capacity regions.
This has been done for Gaussian variants of the \ac{BCC} \cite{Dijk1997, Liu2009} and also by considering
the case of fading channels \cite{Liang2007, Liang2008, Khisti2008, Ekrem2009, Qiao2010}.
More recently, the secrecy capacity regions have been studied for the \ac{BCC} with \ac{MIMO} \cite{Ekrem2012, Liu2013, Bagherikaram2013}
and cooperative communications \cite{Wyrembelski2012}.
In many of these works, coding is considered as an important tool for achieving the reliability and security targets
over the \ac{BCC}, but the abstract model of random coding is often considered \cite{Watanabe2012}, and the design
of practical coding schemes is not addressed.
At the authors' best knowledge, only one proposal of using polar codes as practical codes for transmissions over the 
discrete memoryless \ac{BCC} has very recently appeared \cite{Andersson2013}, while no practical solution exists
for continuous-output \acp{BCC}.

In this work, we consider another important class of powerful error correcting codes, namely \ac{LDPC} codes,
and propose a practical scheme to achieve reliable and secure transmission of public and private information
over the \ac{BCC}.
In order to consider a practically meaningful scenario, our analysis is focused on the case of \ac{QSFC} for both Bob and Eve.
Following some previous literature \cite{Klinc2011, WongWong2011}, the reliability and security performance is measured 
on the basis of the decoding error probabilities experienced at the receiving side.
This way, practical coding schemes can be easily assessed and compared, as we have already done for the 
Gaussian wire-tap channel \cite{Baldi2010, Baldi2011, Baldi2012}.

We show that a transmission scheme able to provide two different levels of protection against noise is needed
to achieve the transmission reliability and security targets over the \ac{BCC}.
For this reason, we use some \ac{LDPC} codes having \ac{UEP} \cite{Poulliat2007, Deetzen2010}.
We consider different modulation formats for the private information bits, and we show that high order
modulations are needed.
% in order to achieve reasonably small outage probabilities over the \ac{BCC} with fading.

The organization of the paper is as follows:
in Section \ref{sec:Model} we define the system model and the metrics we use.
In Section \ref{sec:SysFeas} we address the system feasibility and compute the outage probability for Bob and Eve.
In Section \ref{sec:UEPcodes} we describe the \ac{UEP} \ac{LDPC} codes we propose to use in this context.
In Section \ref{sec:Examples} we provide and discuss some numerical examples, and
Section \ref{sec:Conclusion} concludes the paper.

\section{System model and metrics}
\label{sec:Model}

\begin{figure}[tb]
\begin{centering}
\includegraphics[width=83mm,keepaspectratio]{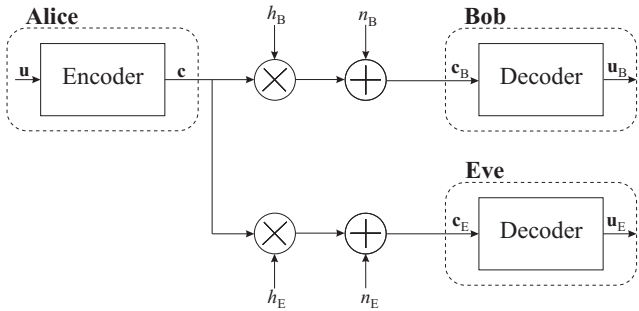}
\caption{Fading wire-tap channel model. \label{fig:wiretap}}
\par\end{centering}
\end{figure}

The channel model we consider is shown in Fig. \ref{fig:wiretap}. 
Both Bob's and Eve's channels are Rayleigh fading, with fading coefficients $h_B$ and $h_E$, respectively, and also affected by \ac{AWGN}, $n_B$ and $n_E$.
We suppose that Bob's and Eve's channels are \acp{QSFC}, that is, their fading coefficients do not vary during the transmission of each codeword,
while they can be modeled as Rayleigh random variables over different codewords.
The \acp{SNR} of Bob's and Eve's channels are usually different. Therefore, the two vectors received by Bob and Eve, $\mathbf{c}_B$ and $\mathbf{c}_E$,
are also different, as well as the two messages they get after decoding, noted by $\mathbf{u}_B$ and $\mathbf{u}_E$, respectively.
Bob is an authorized receiver, able to decode the whole information.
Eve instead is a non-authorized receiver, able to get only the public message information, whereas she should be unable 
to obtain any useful information on the secret message.

Each transmitted message is formed by $n$ bits and includes a public and a confidential part.
Since we use error correcting coding, each transmitted message contains $k$ information bits and $r=n-k$ redundancy bits.
The overall code rate is $R = \frac{k}{n}$, and $R$ also coincides with the overall information rate,
expressed in bits per channel use, when we use \ac{BPSK} modulation.
The transmitted information bits can be divided into a block of $k_s \le k$ secret information bits, 
and another block of $k_p = k - k_s$ public information bits.

The \acp{SNR} on the two channels, noted by $\gB$ and $\gE$, result from the combination of the \ac{AWGN} contribution and the Rayleigh fading contribution.
The average \acp{SNR} are equal to $\gaveB$ and $\gaveE$ for Bob and Eve, respectively.
According to the Rayleigh fading model, $h_B$ and $h_E$ are two Rayleigh random variables, whose
real and imaginary parts are Gaussian random variables with zero mean and variance $1/2$.
Therefore, $|h_B|^2$ and $|h_E|^2$ are chi-square distributed, with average value $E(|h_B|^2) = E(|h_E|^2) = 1$.
It follows that the probability density functions of $\gB$ and $\gE$ are:
\begin{subequations}
\begin{align}
p_{\gB}(x) = \frac{1}{\gaveB} e^{-x/\gaveB},\;\,\;\;\,\,x \geq 0 \label{Pgammap}
\\
p_{\gE}(x) = \frac{1}{\gaveE} e^{-x/\gaveE},\;\,\;\;\,\,x \geq 0 \label{Pgammas}
\end{align}
\label{eq:Pgamma}
\end{subequations}

We suppose to have average \ac{CSI}, that is, Alice knows the values of $\gaveB$ and $\gaveE$.
Several works in the literature assume to have perfect \ac{CSI}, that is, Alice knows 
exactly the values of $\gB$ and $\gE$ for each transmitted codeword.
We prefer to make the assumption of having only average \ac{CSI}, since it is more realistic for
a practical system like the one we want to address.

\subsection{Reliability and security targets}
\label{subsec:Metrics}

In order to design practical coding and modulation schemes for the considered \ac{BCC},
we need some metrics which allow to take into account and assess the performance achieved by
each specific instance of the system.
For this purpose, we adopt the error rate as a metric both for
reliability and security.

Let $P(\gamma)$ denote the overall \ac{FER} as a function of the \ac{SNR} $\gamma$.
In other terms, $P(\gamma)$ is the probability that one or more of the  $k$ information bits 
are in error within a received frame of $n$ bits.
Since each block of $k$ information bits contains a public and a secret part, we denote
by $P_p(\gamma)$ and $P_s(\gamma)$ the \ac{BLER} for each of these two parts, respectively.

As done in some recent literature \cite{Klinc2011, WongWong2011, Baldi2010, Baldi2011, Baldi2012},
we define the security and reliability targets in terms of the decoding error probabilities experienced
by Bob and Eve.
Given two small threshold values, $\delta$ and $\epsilon$, we define the security and reliability targets as follows:
\begin{subequations}
\begin{align}
P_p(\gamma^{(B)}) & \le \delta, \label{eq:Pconda} \\
P_p(\gamma^{(E)}) & \le \delta, \label{eq:Pcondb} \\
P_s(\gamma^{(B)}) & \le \delta, \label{eq:Pcondc} \\
P_s(\gamma^{(E)}) & \ge 1 - \epsilon. \label{eq:Pcondd} 
\end{align}
\label{eq:Pconditions}
\end{subequations}

Conditions \eqref{eq:Pconda}-\eqref{eq:Pcondc} ensure the desired reliability, while \eqref{eq:Pcondd}
guarantees a sufficiently large error probability on the secret information at Eve's.
Having defined the security target in terms of the \ac{BLER}, one could object that, when a block is in error,
this does not necessarily mean that its bits are erred with probability $0.5$ (which would be the desired
maximum uncertainty condition from the information theory standpoint).
Therefore, we cannot state that the system achieves perfect secrecy.
However, we can say that the system achieves a looser notion of \textit{weak secrecy}, as defined in \cite{Bhattad2005}.
In fact, when Eve's \ac{BLER} on the secret information is almost $1$, we know that she has some uncertainty
on the secret information bits.
This small amount of uncertainty can be exploited to achieve a desired higher level of security through suitable
transformations.
For example, an \ac{AONT} \cite{Boyko1999} can be used to link a set of transmitted blocks together,
in such a way that their information can be recovered only when all of them are decoded without errors.
Several examples of \acp{AONT} can be found in the literature.
When transmission occurs over noisy channels, like in this case, we have shown in \cite{Baldi2010, Baldi2011, Baldi2012, Baldi2012a}
that scrambling the information bits through a linear (and dense) map can be sufficient to approach 
an \ac{AONT}, thanks to the randomness of the errors induced by the channel.

On the other hand, using the error rate as a reliability and security metric imposes some restrictions on our analysis.
First of all, the error rate depends on the decoder. Therefore, we should suppose that Eve uses an optimal decoder
to attack the system, that is, a \ac{ML} decoder for continuous-output channels.
However, for sufficiently long \ac{LDPC} codes, it is known that belief propagation iterative decoders are able
to approach the \ac{ML} decoding performance. Therefore, we can consider the performance achieved by Eve
through iterative decoding as a reliable estimate of the optimal decoder performance.
This also avoids the need to consider that Eve uses other decoders, like, for example, those based on \ac{OSD} 
algorithms \cite{Wu2007}, which exploit the concept of \ac{ISD} \cite{Prange1962}, aided by the soft information
available at the channel output.
An algorithm similar to \ac{ISD} was also proposed in \cite{Carrijo2009} to attack two authentication protocols which
exploit some noisy observations.
In that case, however, the attacker can also take advantage of the fact that a fixed secret key is used as the
starting point to compute the transmitted data.

Finally, we observe that the secrecy condition we define by using the error rate as a metric is also weak in 
that it may imply to transmit at a secret rate which is smaller than the equivocation rate at Eve's.
Therefore, there may be some leakage on the confidential information part which must be compensated for
by using higher layer techniques (like \acp{AONT}).
Estimating the equivocation rate of this system and introducing some modifications (like the use of some
intentional randomness) to achieve a secret rate which approaches the equivocation rate will be the
object of future works.

\section{System feasibility and outage}
\label{sec:SysFeas}

\begin{figure}[!t]
\begin{centering}
\includegraphics[width=70mm,keepaspectratio]{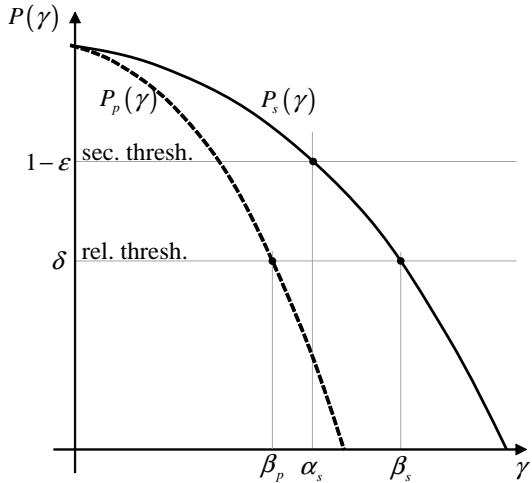}
\caption{Expected block error rate curves for the public and secret messages as functions of the \ac{SNR}.}
\label{fig:ErrorRatePlot}
\par\end{centering}
\end{figure}

Let us suppose that we use a coding and modulation scheme which offers a higher level of protection
against noise to the public information part with respect to the secret information part.
Typical error rate curves for this case are reported in Fig. \ref{fig:ErrorRatePlot}, where
$\beta_p$ and $\beta_s$ denote the minimum \acp{SNR} which are needed to meet the reliability
conditions on the public and the secret information, respectively, while $\alpha_s$ is the maximum 
\ac{SNR} which is allowed to meet the security condition on the secret information.

Based on the $P_p$ and $P_s$ curves, we can rewrite the conditions \eqref{eq:Pconditions} in terms of $\gB$ and $\gE$.
In fact, the conditions \eqref{eq:Pconda} and \eqref{eq:Pcondc} are equivalent to set
\begin{equation}
\gamma^{(B)} \ge \max\left\{\beta_p, \beta_s\right\} = \beta_s,
\label{eq:Bobcond}
\end{equation}
while the conditions \eqref{eq:Pcondb} and \eqref{eq:Pcondd} are satisfied if and only if
\begin{equation}
\beta_p \le \gamma^{(E)} \le \alpha_s.
\label{eq:Evecond}
\end{equation}

Since $1-\epsilon > \delta$ by definition, it follows that the condition \eqref{eq:Evecond} can be met only when the
public information is more protected against noise than the secret information, that is the situation depicted in Fig. \ref{fig:ErrorRatePlot}.
We observe that, in this context, the condition in which Eve has a degraded channel with respect to Bob does not
suffice to make the system feasible as it occurs for the wire-tap channel model.
In principle, the system is feasible even when $\alpha_s = \beta_p$.
In practice, however, we need that $\alpha_s > \beta_p$ to ensure that the system remains feasible even when $\gamma^{(E)}$ 
has some fluctuations, like in the case of fading channels we consider, as we will discuss next.

Provided that the system is feasible, we can assess and compare different coding and modulation schemes by computing the
security gap $S_g$, which is defined as the ratio between the limit values of Bob's and Eve's \acp{SNR} which are needed
to meet the reliability and security conditions.
Designing coding and modulation schemes which achieve small security gaps is important, since this means that reliability and
security can be achieved even with a small degradation of Eve's channel with respect to Bob's channel.

\subsection{Bob's outage}
\label{sec:BobOutage}

When Bob receives a transmitted codeword, he must be able to meet the reliability conditions \eqref{eq:Pconda} and \eqref{eq:Pcondc}.
From \eqref{eq:Bobcond} we have that both these conditions are met when $\gB \ge \beta_s$, hence an outage event occurs when $\gB < \beta_s$.
We denote by $\eta$ the probability of such an event, and from \eqref{Pgammap} we have

\begin{align}
\eta & = P \left\{ 0 \le \gB < \beta_s \right\} = \int_{0}^{\beta_s} p_{\gB}(x) dx \nonumber \\
& = 1 - \exp \left( - \frac{\beta_s}{\gaveB} \right).
\label{eq:eta}
\end{align}

We suppose to have average \ac{CSI} on both channels, hence the transmission power can be chosen
such that the probability of outage is not greater than some fixed value $\eta_{\max}$, that is:

\begin{equation}
\gaveB \ge \gaveB_{\min} = - \frac{\beta_s}{\ln\left(1 - \eta_{\max} \right)}.
\label{eq:gammaBmin}
\end{equation}

\subsection{Eve's outage}
\label{sec:EveOutage}

When Eve receives a transmitted codeword, two outage events can occur:
\begin{itemize}
\item The reliability condition \eqref{eq:Pcondb} on the public information is not met. We define $\omega_r$ the probability of this event.
\item The security condition \eqref{eq:Pcondd} on the secret information is not met. We define $\omega_s$ the probability of this event.
\end{itemize}

Based on \eqref{Pgammas}, we have

\begin{align}
\omega_r & = P \left\{ 0 \le \gE < \beta_p \right\} = \int_{0}^{\beta_p} p_{\gE}(x) dx \nonumber \\
& = 1 - \exp \left( - \frac{\beta_p}{\gaveE} \right)
\label{eq:omegar}
\end{align}
and
\begin{align}
\omega_s & =  P \left\{ \gE > \alpha_s \right\} = \int_{\alpha_s}^{\infty} p_{\gE}(x) dx \nonumber \\
& =  \exp \left( - \frac{\alpha_s}{\gaveE}\right).
\label{eq:omegas}
\end{align}

Since the two outage events are incompatible, the overall outage probability for Eve is
\begin{equation}
\omega = \omega_r + \omega_s = 1 - \exp \left( - \frac{\beta_p}{\overline{\gamma}^{(E)}} \right) + \exp \left( - \frac{\alpha_s}{\overline{\gamma}^{(E)}}\right).
\label{eq:omega}
\end{equation}

As we suppose to have average \ac{CSI} on both channels, we can assume
that $\gaveE$ is chosen in such a way that $\omega$ equals its minimum, $\omega_{\min}$.
This optimal value of $\gaveE$, named $\gaveE_{\mathrm{opt}}$, can be easily found by computing the derivative
of $\omega$ with respect to $\gaveE$, that is,

\begin{equation}
\frac{d\omega}{d\gaveE}=\frac{\alpha_s\exp \left( - \frac{\alpha_s}{\gaveE} \right) - \beta_p\exp \left( - \frac{\beta_p}{\gaveE} \right)}{(\gaveE)^2}.
\label{eq:diff_omega}
\end{equation}
Then, $\gaveE_{\mathrm{opt}}$ is obtained by setting $\frac{d\omega}{d\gaveE}=0 $. This way, we have
\begin{equation}
\gaveE_{\mathrm{opt}} = \frac{\beta_p - \alpha_s}{\mathrm{ln}\left(\frac{\beta_p}{\alpha_s}\right)}.
\label{eq:gamma_min}
\end{equation}

Therefore, by taking Bob's and Eve's outage probabilities (i.e., $\eta_{\max}$ and $\omega_{\min}$) into account,
we can compute the security gap as
\begin{equation}
S_g = \frac{\gaveB_{\min}}{\gaveE_{\mathrm{opt}}}.
\label{eq:Sg}
\end{equation}

\section{\ac{UEP} \ac{LDPC} codes for the \ac{BCC}}
\label{sec:UEPcodes}

In order to achieve the two levels of protection which are requested for
the public and the secret information blocks, we use an \ac{LDPC} code with \ac{UEP}.
We are interested in finding an \ac{UEP} \ac{LDPC} code with a length of $n$ bits,
able to achieve two different levels of protection against noise on the set of $k < n$
information bits.
This requirement fits well with several design approaches proposed in the literature \cite{Poulliat2007, Deetzen2010, Neto2011},
which aim at dividing the codeword bits into three \acp{PC}, named \ac{PC}1, \ac{PC}2 and \ac{PC}3, respectively:

\begin{itemize}
\item The \ac{PC}1 contains $k_1 < k$ information bits which are the most protected against noise.
\item The \ac{PC}2 contains $k_2 = k - k_1$ information bits which are less protected against noise than those in \ac{PC}1.
\item The \ac{PC}3 contains the $r = n-k$ redundancy bits.
\end{itemize}

Hence, for the use in the considered scenario, we can design a code with these three \acp{PC},
and then map the public information bits into \ac{PC}1 (i.e., $k_p = k_1$) and the secret information
bits into \ac{PC}2 (i.e., $k_s = k_2$). 

Codes of this kind can be obtained by designing suitable node degree distributions and then
grouping the codeword bits based on their node degrees.
More in detail, first of all, the variable node degree distribution must be chosen in such a way as to achieve a good convergence
threshold under iterative decoding. To have \ac{UEP}, instead, the degree distribution must include both very low and rather 
high variable node degrees. Since the highest degrees ensure greater protection, the variable nodes with such degrees which correspond to information bits form the \ac{PC}1. Among the remaining variable nodes, with low degrees, those corresponding to information bits 
form the \ac{PC}2. Finally, the variable nodes associated to redundancy bits form the \ac{PC}3.

The design starts from an optimized variable node degree distribution from the edge perspective, 
which is expressed as a polynomial, $\lambda(x) = \sum _{i=1}^{\overline{d_{v} }}\lambda _{i} x^{i-1}$, with real coefficients.
The coefficient $\lambda _{i}$ coincides with the fraction of edges connected to variable nodes having degree $i$,
and $\overline{d_{v}}$ is the maximum variable node degree.
Then, $\lambda(x)$ is converted from the edge perspective to the node perspective, thus obtaining 
the polynomial $\nu(x) = \sum_{i=1}^{\overline{d_{v} }} \nu_i x^i$, whose coefficients $\nu_i$ are related to the $\lambda_i$'s as follows:
\begin{align}
\nu_i  & = \frac{\lambda_i/ i}{\sum_{j=1}^{\overline{d_v}} \lambda_j/j}, \nonumber \\
\lambda_i  & = \frac{\nu_i \cdot i}{\sum_{j=1}^{\overline{d_v}} \nu_j \cdot j}.
\end{align}

The node perspective is useful to assign the variable node bits to the \acp{PC}. In particular, the number of bits in \ac{PC}1, that is the most protected class, is computed by summing the fractions of nodes (\textit{i.e.}, the values of $\nu_i$) corresponding
to the highest values of $i$ (\textit{i.e.}, to the highest variable node degrees).

The same reasoning can be applied to the check nodes degree distributions, by denoting with $\rho(x)$ and $c(x)$
the check node degree distributions from the edge and the node perspectives, respectively, and by replacing $\lambda$ with $\rho$,
$\nu$ with $c$, and $\overline{d_v}$ with $\overline{d_c}$, where $\overline{d_c}$ is the maximum check node degree.

Concerning the design of the check node degree distribution, we adopt a concentrated distribution (i.e., with only two degrees, concentrated around the mean).
This solution has the advantage of being very simple, while achieving good performance. This way, we obtain
\begin{equation}
c(x) = a x^{\left\lfloor c_m \right\rfloor} + b x^{\left\lceil c_m \right\rceil},
\end{equation}
where $c_m = \frac{E}{r} = \frac{\sum_{j} v_j \cdot j}{(1-R)}$ and $E$ is the total number 
of edges in the Tanner graph.
The values $a$ and $b$ are computed as
\begin{equation}
a = \lceil c_m \rceil - c_m, \ \ \ \ b = c_m - \lfloor c_m \rfloor.
\end{equation}

Once having designed the variable and check nodes degree distributions, a practical code with an arbitrary finite length
can be obtained by designing its $r \times n$ parity-check matrix in such a way as to match the two degree
distributions.
This can be accomplished through several algorithms. Among them, we adopt the \textit{zigzag-random} construction \cite{Hu2001PEG, Deetzen2010}.

Since we use these codes to map the first two \acp{PC} to the public and the secret information bits,
it is advisable to ensure that a high level of separation exists between these two classes, in such a way that possible 
fluctuations of the error rate on one of them do not affect the error rate on the other.
For this purpose, we design the parity-check matrix in such a way as to keep the number of parity-check equations
which are common between the first two \acp{PC} as small as possible, while still achieving good performance.

\section{Numerical examples}
\label{sec:Examples}

In order to provide some numerical examples, we consider an \ac{UEP} \ac{LDPC} code
with $n=4096$ and overall code rate $R = 1/2$.
Its variable nodes degree distribution is taken from \cite[Table 3]{Poulliat2007}, with some 
minor modifications which are needed to change the proportion between the \ac{PC}1 and the \ac{PC}2:
\begin{align}
\lambda (x) & = 0.0025 x^{19} + 0.0009 x^{18} + 0.0031 x^{17} + 0.0630 x^{16} + \nonumber \\
                & + 0.3893 x^{15} + 0.2985 x^{2} + 0.2427 x.
\end{align}
The corresponding degree distribution from the node perspective is
\begin{align}
\nu (x) & = 0.0005 x^{20} + 0.0002 x^{19} + 0.0007 x^{18} + 0.0151 x^{17} + \nonumber \\
          & + 0.0835 x^{16} + 0.4054 x^{3} + 0.4946 x^{2}.
\label{eq:nu}
\end{align}

We observe from \eqref{eq:nu} that the variable nodes can be grouped into two classes with
node degrees $\le 3$ or $\ge 16$.
Therefore, the nodes in \ac{PC}1 will be those having degree $\ge 16$, while the others will
be in \ac{PC}2 or \ac{PC}3, depending on their association to information or redundancy bits.
This way, we find that \ac{PC}1 and \ac{PC}2
contain, respectively, $20\%$ and $80\%$ of the information bits.

\begin{figure}[!t]
\begin{centering}
\includegraphics[width=80mm,keepaspectratio]{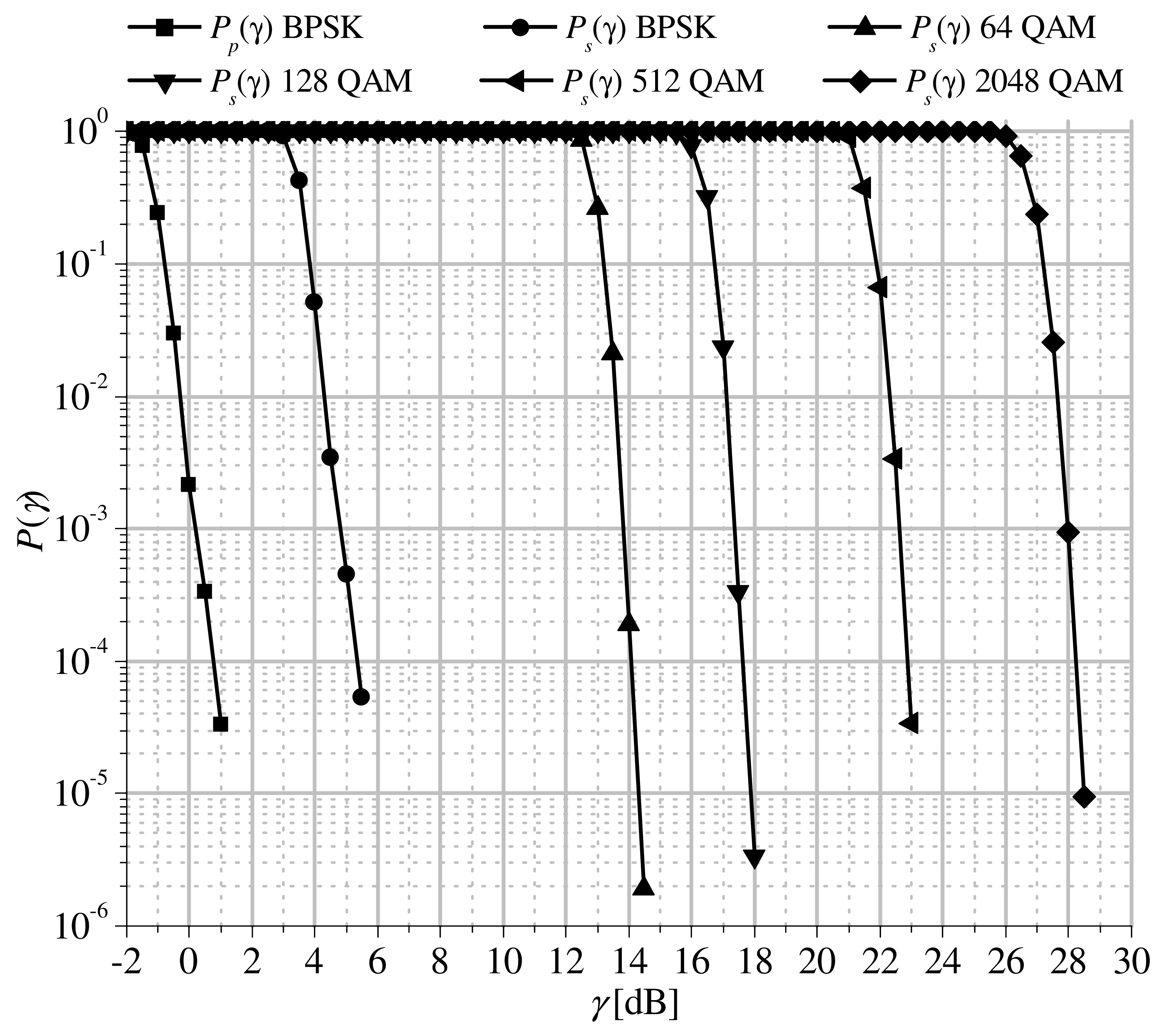}
\caption{Error rate curves for an \ac{UEP} \ac{LDPC} code with length $4096$ and \ac{PC}1 and \ac{PC}2 with proportions $20\%-80\%$.
The bits in the \ac{PC}1 are always BPSK modulated, while the performance of several \ac{QAM} schemes with Yarg labeling on the bits in the \ac{PC}2 is reported.}
\label{fig:UEP4096}
\par\end{centering}
\end{figure}

The performance of this code has been assessed by simulating transmission over a Gaussian
channel with \ac{SNR} per bit equal to $\gamma$, and by performing decoding through the
\ac{LLR-SPA}.
The bits in the \ac{PC}1 are always transmitted by using \ac{BPSK} modulation, while for the
bits in the \ac{PC}2 several \ac{QAM} formats have also been tested.
For the latter, we have adopted the labeling known as Yarg \cite{Kwak2009}, which has
been suitably designed for physical layer security contexts.
Concerning \ac{QAM} transmissions, they have been implemented through a pragmatic
approach, by mapping groups of bits into \ac{QAM} symbols, and then using a classical
symbol-to-bit soft metric conversion before \ac{LDPC} decoding.
The performance obtained, in terms of $P_p(\gamma)$ and $P_s(\gamma)$, is reported
in Fig. \ref{fig:UEP4096}.

We fix two values for the reliability and security thresholds, that is, $\delta = 10^{-4}$ and $\epsilon = 0.1$.
Based on these choices, from Fig. \ref{fig:UEP4096} we obtain $\beta_p=0.75$ dB, while $\alpha_s$ and $\beta_s$ vary 
according to the modulation scheme used for the secret information bits.
The values taken by $\alpha_s$ and $\beta_s$ for the considered modulation schemes are reported in Table \ref{tab:Comparison}.

\begin{table}[!t]
\renewcommand{\arraystretch}{1.1}
\caption{Performance of the considered coding and modulation schemes (all values are in dB, except the outage probability)}
\label{tab:Comparison}
\centering
\begin{tabular}{c|c|c|c|c|c|c}
\hline
\multirow{2}{*}{Scheme} & \multirow{2}{*}{$\alpha_s$} & $\omega_{\min}$ & \multirow{2}{*}{$\gaveE_{\mathrm{opt}}$} & \multirow{2}{*}{$\beta_s$} & \multirow{2}{*}{$\gaveB_{\min}$} & \multirow{2}{*}{$S_g$} \\
	         &  & $(\eta_{\max})$ & & & & \\
\hline
\hline
BPSK & $2.95$ & $0.81$ & $1.90$ & $5.35$ & $3.14$ & $1.24$\\
\hline
64 QAM & $12.25$ & $0.24$ & $7.70$ & $14.12$ & $19.73$ & $12.03$\\
\hline
128  QAM & $15.78$ & $0.13$ & $10.25$ & $17.67$ & $26.23$ & $15.98$\\
\hline
512 QAM & $20.64$ & $0.05$ & $13.99$ & $22.94$ & $35.84$ & $21.85$\\
\hline
%1024 QAM & $13.02$ & $0.26$ & $8.62$ & $17.02$ & $26.89$ & $18.27$\\
%\hline
2048 QAM & $25.27$ & $0.02$ & $17.73$ & $28.49$ & $45.44$ & $27.71$\\
\hline
\end{tabular}
\end{table}

\begin{figure}[!t]
\begin{centering}
\includegraphics[width=80mm,keepaspectratio]{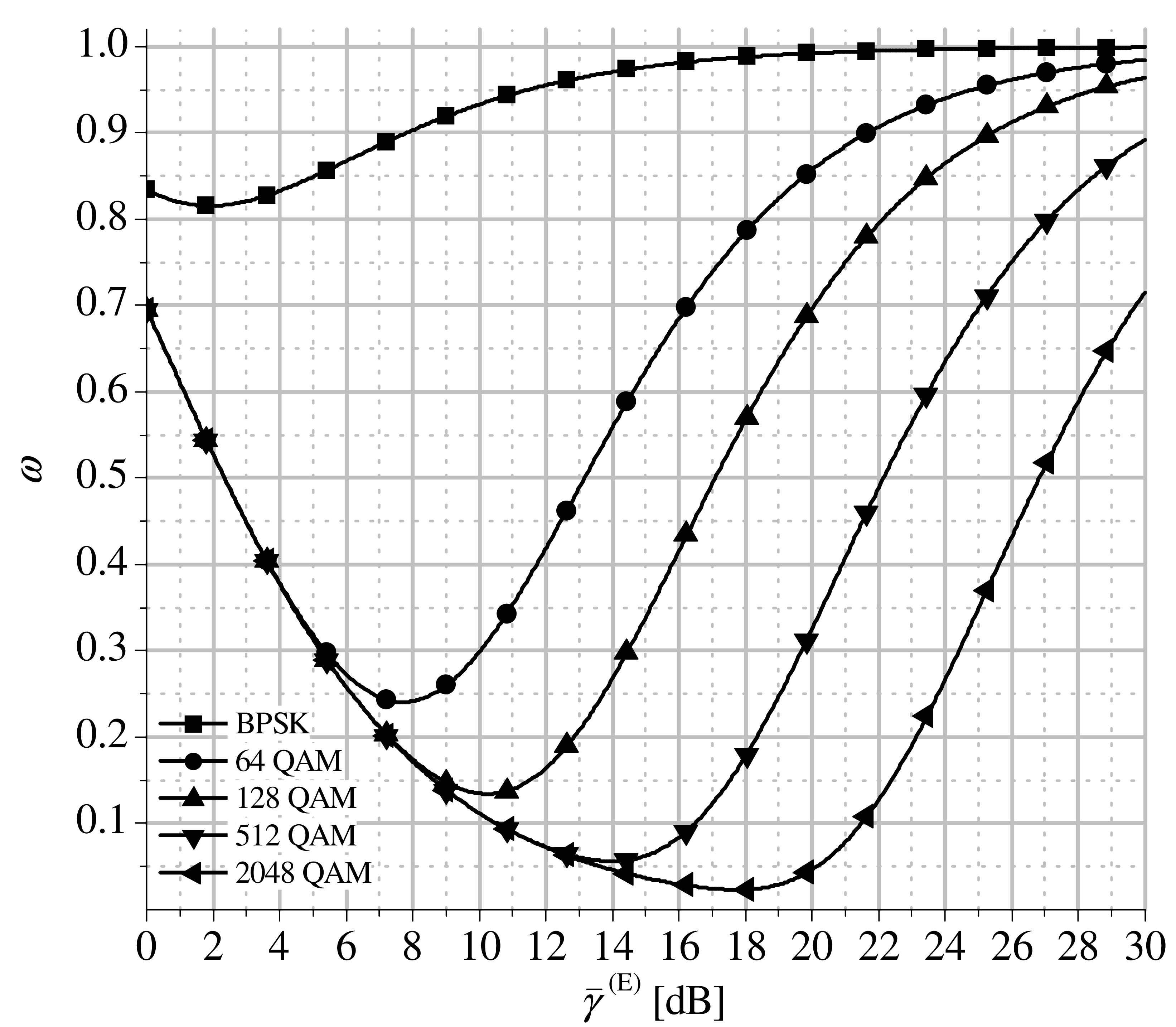}
\caption{Eve's outage probability $\omega$ as a function of Eve's average \ac{SNR} per bit $\gaveE$ for an \ac{UEP} \ac{LDPC}
coded transmission with BPSK-modulated public information bits and several \ac{QAM} formats with Yarg labeling on the secret information bits.}
\label{fig:omega}
\par\end{centering}
\end{figure}

Starting from the values of $\alpha_s$ and $\beta_s$, we can compute Eve's overall outage probability $\omega$ \eqref{eq:omega},
as a function of Eve's average \ac{SNR} per bit, $\gaveE$. The values of $\omega$, so obtained, are reported in Fig. \ref{fig:omega}
for the considered secret information modulation formats.
Then, the value of $\omega_{\min}$ is easily obtained, as well as the value of $\gaveE_{\mathrm{opt}}$ for which $\omega = \omega_{\min}$.
These values can be found according to the procedure described in Section \ref{sec:EveOutage}, and are also reported in Table \ref{tab:Comparison}.
Concerning Bob, we have fixed a maximum outage probability $\eta_{\max} = \omega_{\min}$, and computed the corresponding minimum value 
of his average \ac{SNR} per bit, $\gaveB_{\min}$, according to \eqref{eq:gammaBmin}. The values of $\gaveB_{\min}$, so obtained, are also 
reported in Table \ref{tab:Comparison}.

Based on these results, we observe that, when both the public and the secret information bits are modulated with BPSK,
the outage probability for Eve is always very large (more than $0.8$). Therefore, although the system is theoretically feasible, in practice the
fading nature of the channel rarely allows to achieve a successful transmission.
The situation improves by adopting higher modulation orders for the private information bits, which also increases the
values of $\alpha_s$. This way, the outage probability for Eve is progressively reduced.
When we adopt a \ac{QAM} scheme with $2048$ symbols, Eve's outage probability can be reduced down to $0.02$.
Under the hypothesis that Bob's outage probability is the same as Eve's outage probability (or less), we observe that
there is a tradeoff between the outage probability and the security gap.
In fact, if we are able to tolerate a high probability of outage, the system requires small security gaps (in the
order of 10 dBs or even less).
Instead, if we aim at small outage probabilities, we need large security gaps (in the order of 20 or 30 dBs).

On the other hand, high values of the outage probability (which mean that the system is often in outage) may be not much appealing in practice.
However, as often occurs in physical layer security, the proposed coding and modulation scheme is able to offer a basic level of reliability and security,
which can be then exploited by higher layer protocols and algorithms to reach the desired performance.
Some examples of higher layer techniques of this kind are automatic repeat request protocols, to improve the transmission reliability,
and \acp{AONT} acting on groups of concatenated blocks, to improve the transmission security without the need of any shared secret.

% Actually, the goal of this paper was to demonstrate the feasibility of the system. Working on the degrees of freedom and further optimizing the code design, substantial improvements are expected.

\begin{comment}
By looking at these outage values, one could object that the usefulness of this system is questionable, since it is
in outage for about one transmission each ten.
However, we remind that the aim of this transmission scheme is to offer a basic level of secrecy, while any
desired level of security can be achieved by using further tools, like \acp{AONT} acting on groups of concatenated blocks.
\end{comment}

\section{Conclusion}

We have studied the \ac{BCC} with quasi-static fading from a practical standpoint, by using the decoding error
probability as a metric.
We have proposed some practical coding and modulation schemes which can achieve some fixed reliability and 
security targets over this channel.

We have computed closed form expressions for the probability of outage at Bob's and Eve's, and assessed
the security gap which is needed between their channels under the hypothesis of average \ac{CSI}.

Our results show that high order modulation schemes are advisable for the secret information bits in order
to achieve reasonably low values of the outage probability, although this yields some increase in the
security gap.

As anticipated in Section \ref{subsec:Metrics}, future works will involve the assessment of the performance
achievable in terms of the equivocation rate at Eve's, and the optimization of the coding and modulation scheme
to work at a secret rate which approaches the equivocation rate.

\label{sec:Conclusion}

\newcommand{\BIBdecl}{\setlength{\itemsep}{0.005\baselineskip}}
\bibliographystyle{IEEEtran}
\bibliography{Archive}

\end{document}